\documentclass[11pt]{article}
\usepackage[left=0.8in,right=0.8in,a4paper]{geometry}
\usepackage{empheq}
\usepackage{amssymb}
\usepackage[pdfstartview=FitH,pdfpagemode=None]{hyperref}
\usepackage{color}
\usepackage{slashed}
\usepackage{amsmath}
\usepackage{amsfonts}
\usepackage{amssymb}
\usepackage{authblk}
\usepackage{graphicx}
\usepackage{caption}
\usepackage{subcaption}
\usepackage{float}
\usepackage{appendix}
\usepackage{listings}

\newcommand{\be}{\begin{equation}}
\newcommand{\ee}{\end{equation}}
\newcommand{\ba}{\begin{eqnarray}}
\newcommand{\ea}{\end{eqnarray}}

\newcommand{\tr}{\text{tr}}
\begin{document}
\title{Circular Wilson loops in defect Conformal Field Theory}
\author[a]{Jerem\'{\i}as Aguilera-Damia\thanks{jeremiasadlp@gmail.com}}
\author[a]{Diego H. Correa \thanks{correa@fisica.unlp.edu.ar}}
\author[b]{Victor I. Giraldo-Rivera\thanks{vgiraldo@icts.res.in}}

\date{}

\affil[a]{ Instituto de F\'isica La Plata, CONICET,
Universidad Nacional de La Plata
C.C. 67, 1900 La Plata, Argentina}
\affil[b]{International Centre for Theoretical Sciences (ICTS-TIFR),
Shivakote, Hesaraghatta Hobli, Bengaluru 560089, India.}
\maketitle

\begin{abstract}
We study a D3-D5 system dual to a conformal field theory with a codimension-one defect that separates regions where the ranks of the gauge groups differ by $k$. With the help of this additional parameter, as observed by Nagasaki, Tanida and Yamaguchi, one can define a double scaling limit in which the quantum corrections are organized in powers of $\lambda/k^2$, which should allow to extrapolate results between weak and strong coupling regimes. In particular we consider a radius $R$ circular Wilson loop placed at a distance $L$, whose internal space orientation is given by an angle $\chi$. We compute its vacuum expectation value and show that, in the double scaling limit and for small $\chi$ and small $L/R$, weak coupling results can be extrapolated to the strong coupling limit.
\end{abstract}


\section{Introduction}
\label{intro}

The study of intersecting D3-D$p$ branes has led to interesting realizations of conformal field theories in the context of the AdS/CFT correspondence. In the near horizon limit, a single D$p$-brane is seen as a probe brane in AdS$_5\times$S$^5$. We will be concerned in realizations in which the dual description leads to a defect or domain wall that separates an ${\cal N} = 4$ super Yang-Mills with gauge group SU$(N)$ from another one with gauge group SU$(N-k)$.

In the case of a general D3-D5 intersection, the additional defect has codimension one and is set, for definiteness, at $x_3 =0$. When $k$ D3-branes out of the stack of $N$ terminate  on a stack of $M$ D5-branes, the dual gauge group is SU$(N - k)$ for $x_3 > 0$, while it is SU$(N)$ for $x_3 < 0$ \cite{Karch:2000gx}. The dual gauge theory introduces $M$ fundamental hypermutiplets living on the 3D defect and interacting with the usual $\mathcal{N}=4$ vector multiplet field content \cite{DeWolfe:2001pq}, being superconformal for general $M$. However, in the near horizon limit, realization in terms of M probe D5-branes on $AdS_5\times S^5$ is only valid for $M\ll N$. In particular, the realization we are interested in this article involves only one of such branes, which implies $M=1$. The original supersymmetry is broken to $\text{OSp}(4|4)\subset \text{PSU}(2,2|4)$ and consequently the $\mathcal{N}=4$ vector multiplet splits on a vector and a hypermultiplet in 3D with the corresponding R-symmetry breaking $\text{SU}(4)\rightarrow \text{SO}(3)_V\times \text{SO}(3)_H$.

This D3-D$p$ brane constructions extended the landscape for generalizing the full set of techniques developed in previous realizations. In particular, a state-operator correspondence was established in the BMN limit \cite{Lee:2002cu} and the one-loop dilatation operator was mapped to an integrable spin chain in the scalar sector \cite{DeWolfe:2004zt}. Moreover, integrability of these realizations in both gauge and string theory side was intensively studied by constructing the corresponding Bethe system and solving for open string configurations attached to the D5-brane \cite{Okamura:2005cj,Susaki:2005qn,Susaki:2004tg,chen}.

Also a novel feature which is inherent to these new realizations has been object of several works in the last few years, namely that gauge symmetry breaking at one side of the defect is induced by $k$ components of the scalar fields acquiring non-zero vacuum expectation values  \cite{Diaconescu:1996rk,Giveon:1998sr,Constable:1999ac}. Moreover, there is a prescription for computing these objects on the gravity side. Vacuum expectation values for this set of operators were studied in both weak and strong coupling regimes for either non-supersymmetric D3-D7  \cite{Kristjansen:2012tn} and supersymmetric D3-D5 realizations \cite{Kristonepoint, Buhl-Mortensen:2016jqo}. Furthermore, one-point functions for non BPS single trace operators have been worked out in terms of integrable spin chains \cite{deLeeuw:2015hxa, Buhl-Mortensen:2015gfd,deLeeuw:2016umh}.

Following \cite{KristWL,Nagasaki,Nagasaki:2012re}, a double scaling limit can be considered for those defect conformal field theories leading to a remarkable feature. Gravity computations, which valid for large 't Hooft coupling $\lambda$, can be considered for large $k$ in such a way that $\lambda/k^2$ is kept small and the results are found to be expressible in powers of $\lambda/k^2$. Thus, in that regime, it is possible to successfully compare gauge and gravity results providing further non-trivial verifications of the AdS/CFT correspondence.

Our goal is to study Wilson loops in this context and in particular their expectation values in the double scaling limit which allows to compare perturbative with string theory results. Computations of Wilson loop operators in the presence of defects were first considered in \cite{KristWL,Nagasaki}. In particular we will consider circular Wilson loops, analogue to the supersymmetric ones in ordinary ${\cal N} = 4$ super Yang-Mills which could be studied by means of localization techniques \cite{Pestun:2007rz}.

We would like to compute the vacuum expectation value of a circular Wilson loop of radius $R$ placed at a distance $L$ from the defect. We shall consider the following Euclidean Wilson loop
\begin{align}
W &=\text{tr}\mathcal{P} \exp\left\{\oint d\tau
\left[iA_\mu \dot x^\mu -|\dot x|(\sin\chi \Phi_3 +\cos\chi \Phi_6)\right]\right\}\,,
\label{wl}
\end{align}
where $\chi$ is taken to be some parameter on the interval $[0,\frac{\pi}{2}]$. If we parametrize
the circle as
\be
x^{\mu}(\tau)=\left(0,R\cos\tau,R\sin\tau,L\right)\,,
\ee
we get
\begin{align}
W &=\text{tr}\mathcal{P} \exp\left\{R\int_0^{2\pi} d\tau
\left[-i A_1 \sin\tau + i A_2 \cos\tau -\sin\chi \Phi_3 -\cos\chi \Phi_6\right]\right\}\,.
\label{wl2}
\end{align}

Note that by conformal invariance $\langle W \rangle$ depends on $R$ and $L$ only through the ratio $R/L$. So that, the expectation value $\langle W \rangle$ depends on the parameters of the gauge theory $\lambda$, $N$ and $k$ as well as on the parameters $R/L$ and $\chi$ of the Wilson loop.  We will explore different regimes for all these parameters, in the weak coupling limit through perturbative computations and in the strong coupling limit through string theory computations. We will consider the extrapolation of weak coupling results to the strong coupling regime in the double scaling limit, for the case of small $\chi$ and small $L/R$. We will also analyze what are the requirements for the Wilson loop to be supersymmetric. Requirements for this operator to preserve some amount of the supercharges preserved by de interface are analysed in appendix \ref{susy}, where condition $\chi=0$ has been found.

\section{Classical string dual the circular Wilson loop}\label{string}

The holographic representation of the theory consist on type IIB string theory in AdS$_5\times$S$^5$ background with a D5-brane ending at the position of the defect ({\it i.e.} $x_3=0$) at the boundary. Such a brane configuration corresponds to a solution of the DBI action extended along AdS$_4\times$S$^2$ with $\kappa=\frac{\pi k}{\sqrt{\lambda}}$ units of magnetic flux. With this definition, in the double scaling limit when $\tfrac{\lambda}{k^2}$ is kept fixed and small, we have to to keep $\kappa$ fixed and large.

We will take the AdS metric in the Poincar\'e patch
\be
ds^2_{\rm AdS}=\frac{1}{y^2}\left(-dt^2+dy^2+dr^2+r^2d\phi^2+dx^2_3\right)\,,
\ee
and for the sphere
\be
ds^2_{{\rm S}^5}=d\theta^2+\sin^2\theta d\Omega_2^2+\cos^2\theta d\tilde\Omega_2^2\,,
\ee
where $\Omega_2$ and $\tilde\Omega_2$ denote two $S^2$ spheres.
In these coordinates the D5-brane solution is
\be
y=\frac{1}{\kappa}x_3\,, \qquad
 \mathcal{F}=-\kappa \text{Vol} (S^2)\,, \qquad
\theta=\frac{\pi}{2}\,.
\ee

In what follows, we will consider a fundamental string stretching from the boundary to the D5-brane. For the classical string to be dual to the circular Wilson loop we will impose that at the boundary the string worldsheet terminates at $x_3 =L$ on a circle of radius $R$.

We propose the following ansatz
\be
y=y(\sigma)\,, \qquad r=r(\sigma)\,, \qquad \phi=\tau\,,\qquad x_3=x_3(\sigma)\,,\qquad  \theta=\theta(\sigma)\,.
\ee
Then, the Polyakov action in the conformal gauge reads
\be
S=\frac{\sqrt{\lambda}}{4\pi}\int d\tau d\sigma
\frac{1}{y^2}\left(y'^2+r'^2+r^2+x'^2_3 +y^2\theta'^2\right)\,,
\ee
and the Virasoro constraint becomes
\be
y'^2+r'^2+x'^2_3 +y^2\theta'^2=r^2\,.
\label{vir}
\ee
The equations of motion for $x_3$ and $\theta$ introduce two constants of motion
\be
x'_3=-c y^2\,,  \qquad  \theta'= m\,,
\label{cm}
\ee
and the equations for $y(\sigma)$ and $r(\sigma)$ become
\be
y y''+r'^2+r^2-y'^2+c^2y^4=0\,, \qquad  y r'' -2r'y'-y r=0\,.
\label{eom}
\ee
In order for the string to end on the D5-brane, solutions of \eqref{eom} are subject to the following boundary conditions at the D5-brane
\begin{align}
y'(\tilde{\sigma})-\kappa cy^2(\tilde{\sigma})&=0\,,\qquad
 r' (\tilde{\sigma}) =0\,,\nonumber\\
y(\tilde{\sigma})-\frac{1}{\kappa}x_3(\tilde{\sigma})&=0\,,\qquad
\theta(\tilde{\sigma})=\frac{\pi}{2}\,,
\label{bc}
\end{align}
where $\tilde{\sigma}$ denotes the maximum value of the $\sigma$-variable. On the other hand, conditions at the AdS boundary, achieved for $\sigma\to 0$, are
\begin{align}
y(0)&=0\,,\qquad r(0)=R\,,\nonumber\\
x_3(0)&=L\,,\qquad \theta(0)=\chi\,.
\label{bound}
\end{align}

The solution for  $\theta$ is, by means of \eqref{cm}
\be
\theta(\sigma)= m \sigma + \chi\,,
\label{theta}
\ee
where $\chi\in[0,\frac{\pi}{2}]$, the value of $\theta$ at the boundary, is in correspondence with the parameter $\chi$ of the Wilson loop (\ref{wl2}).

For general $c$ and $m$ finding an exact solution results in a hard task. We will begin by presenting a solution for $c=0$ and then expand around it.

\subsection{Solution for $c=0$}

In this limit $\chi$ will not be an independent parameter anymore. Moreover it will depend in a non-trivial way on $m$. Eventually, we will be interested in the large $k$ limit, which requires large $m$ and $\chi\to 0$. We can establish a non trivial comparison with gauge theory calculations even in this limit.
For $c=0$, $x_3$ is constant and decouple from the equations of motion for $y(\sigma)$ and $r(\sigma)$ which read\footnote{We have used the Virasoro constrain
in the equation for $y(\sigma)$.}
\be
y y'' + 2\left(r'\right)^2 + m^2 y^2=0\,, \qquad y r'' -2 y' r' - y r =0\,.
\label{eomsc0}
\ee
For later convenience we define a new variable
\be
x=\sqrt{1+m^2}\sigma\,,
\label{change}
\ee
and the equations (\ref{eomsc0}) become
\be
y y'' + 2\left(r'\right)^2 + \frac{m^2}{1+m^2} y^2 =0\,,\qquad
y r'' -2 y' r' - \frac{yr}{1+m^2}=0\,,
\ee
where now $'$ stands for derivatives with respect to $x$. Solutions to these equations satisfying boundary conditions \eqref{bound} can has been found and expressed in terms of Jacobi elliptic functions,
\begin{align}
y(x)= y_0(x)&=\frac{R}{\sqrt{1+m^2}} \;\text{sn}\left(x,\tfrac{1}{1+m^2}\right)\,,
\\  r(x)= r_0(x)&= R \;\text{dn} \left(x,\tfrac{1}{1+m^2}\right)\,.
\label{solution}
\end{align}
It is easy to see that $y^2+r^2=R^2$ and that it satisfies Virasoro constraint \eqref{vir}. The first boundary condition in \eqref{bc} imposes
\be
\text{cn}\left(\tilde{x}_0,\tfrac{1}{1+m^2}\right)\;\text{dn}\left(\tilde{x}_0,\tfrac{1}{1+m^2}\right)=0 \,,
\ee
which relates $\tilde{x}_0=\sqrt{1+m^2}\tilde{\sigma}_0$, the maximum value of the $x$-variable, with $m$.
Both cn and dn are bilocal functions and their zeroes are of the form $(2n+1)\mathbb{K}\left(\tfrac{1}{1+m^2}\right)+i2n'\mathbb{K}\left(\tfrac{m^2}{1+m^2}\right)$ and $(2n+1)\mathbb{K}\left(\tfrac{1}{1+m^2}\right)+i(2n'+1)\mathbb{K}\left(\tfrac{m^2}{1+m^2}\right)$ respectively, where $\mathbb{K}$ denotes the complete elliptic integral of the first kind and $n,n'\in\mathbb{Z}$. The minimum real zero occurs for $n=n'=0$, thus we obtain
\be
\tilde{x}_0=\mathbb{K}\left(\tfrac{1}{1+m^2}\right)\,.
\label{sigma0}
\ee
The last equation from \eqref{bc} tells that the parameter $\chi$ is related to $m$ as well,
\be
\chi=\frac{\pi}{2}-m\tilde{\sigma}_0 =
\frac{\pi}{2}-\frac{m}{\sqrt{1+m^2}}\mathbb{K}\left(\tfrac{1}{1+m^2}\right)
\label{chi0}
\ee
Since we are eventually interested in the large $\kappa$ limit we should know the relation between $\kappa$
and $m$. This is obtained from the third equation in \eqref{bc} that gives
\be
m = \sqrt{\left(\frac{\kappa R}{L}\right)^2-1}\,.
\label{m0}
\ee

In order to evaluate the action on-shell we must regularize it by introducing a cutoff $\epsilon$ in the lower integration limit for $\sigma$. The regularized action becomes
\begin{align}
S_0 =\frac{\sqrt{\lambda}}{\sqrt{1+m^2}}\int^{\tilde{x}_0}_{\text{reg}}dx
\frac{r_0^2}{y_0^2}&=\frac{\sqrt{\lambda}}{1+m^2}
\left(m^2\mathbb{K}\left(\tfrac{1}{1+m^2}\right)-(1+m^2)\mathbb{E}\left(\tfrac{1}{1+m^2}\right)\right)\nonumber
\\
&=\tfrac{\pi k R}{L}\left(1-\tfrac{\lambda L^2}{\pi^2 k^2 R^2}\right)\mathbb{K}(\tfrac{\lambda L^2}{\pi^2 k^2 R^2})
-\tfrac{\pi k R}{L}\mathbb{E}(\tfrac{\lambda L^2}{\pi^2 k^2 R^2})\,,
\label{exact}
\end{align}
where $\mathbb{E}$ denotes the complete elliptic integral of the second kind.  For large $\kappa =\tfrac{\pi k}{\sqrt{\lambda}}$ we get an expansion in powers of $\tfrac{\lambda L^2}{k^2 R^2}$
\be
S_0 = -\frac{k R \pi^2}{L}\left(\frac{1}{4}\frac{\lambda L^2}{\pi^2 k^2 R^2}+
\frac{1}{32}\frac{\lambda^2 L^4}{\pi^4 k^4 R^4}+\frac{3}{256}\frac{\lambda^3 L^6}{\pi^6 k^6 R^6}
+\mathcal{O}\left(\tfrac{\lambda L^2}{k^2 R^2}\right)^4\right)\,.
\label{prediction}
\ee
The effective parameter of this expansion  can be small even if $\lambda$ is large, provided $\tfrac{k^2 R^2}{L^2}$ is much larger. As we will see in section \ref{perturbative}, the gauge theory perturbative computation of the Wilson loop expectation value will be also naturally organized in powers of $\tfrac{\lambda L^2}{k^2 R^2}$. Therefore, our on-shell action (\ref{prediction}) is a prediction for the successive loop orders for the expectation values of a Wilson loop of radius $R$, at a distance $L$ of the defect and with $\chi$ given by
\be
\chi=
\frac{\pi}{2} -\sqrt{1-\tfrac{\lambda L^2}{\pi^2 k^2 R^2}}\mathbb{K}(\tfrac{\lambda L^2}{\pi^2 k^2 R^2})
=
\left(\frac{1}{8}\frac{\lambda L^2}{k^2 R^2 \pi^2}+
\frac{7}{128}\frac{\lambda^2 L^4}{k^4 R^4 \pi^4}
+\mathcal{O}\left(\tfrac{\lambda L^2}{k^2 R^2}\right)^3\right)\,.
\label{chi0bisbs}
\ee
We will verify the first term in the expansion (\ref{prediction}) with a 1-loop perturbative computation.

\subsection{Solution for $c\neq 0$}

In the previous subsection we have found an expansion in powers of $\lambda/k^2$ for the expectation value of a circular the Wilson loop coupled in internal space with a very specific angle $\chi$ \eqref{chi0bisbs}. Finding a solution corrected by powers of the parameter $c$, will be obviously a more interesting setting. Moreover, we will later show that \eqref{chi0bisbs} does not correspond to any supersymmetric configuration, which is another motivation for looking configurations with more generic values of $\chi$.

However, finding an exact solution for arbitrary $c$ and $m$ is difficult, so we propose an small $c$ expansion of the
form\footnote{It is convenient to change $c = \tilde{c}\sqrt{1+m^2}$ as well. From now on, the expansion will be in powers of $\tilde{c}$ but we will
omit $\tilde{\phantom{c}}$ in the the notation.}
\be
\begin{array}{rcl}
y(x)&=&y_0(x)+c y_1(x)+c^2 y_2(x) + {\cal O}(c^3)\,,\\
r(x)&=&r_0(x)+c r_1(x)+c^2 r_2(x) + {\cal O}(c^3)\,,
\label{expansion}
\end{array}
\ee
where $y_0$ and $r_0$ were defined in \eqref{solution}.

On the other hand, parameters $\tilde{x}$, $\kappa$ entering in the boundary conditions \eqref{bc} will be functions of $c$ and $m$ as well. So we consider the
following expansions for them
\begin{align}
\tilde{x} = \tilde{x}_0 + c \tilde{x}_1 + c^2 \tilde{x}_2 + {\cal O}(c^3)\,,
\label{xexp}
\\
\kappa  = \kappa_0 + c \kappa_1 + c^2 \kappa_2 + {\cal O}(c^3)\,,
\label{kapexp}
\end{align}
where each $\tilde{x}_a$ and $\kappa_a$ are functions of $m$. 
Parameter $\chi$ is also a function of $c$ and $m$ through
\be
\chi=\frac{\pi}{2}-\frac{m}{\sqrt{1+m^2}}\left(\tilde{x}_0 + c \tilde{x}_1 + c^2 \tilde{x}_2\right) + {\cal O}(c^3) \,.
\label{chiexp}
\ee

Eventually, we would like to trade parameters $c$ and $m$ by parameters $\kappa$ and $\chi$ which is achieved by inverting relations (\ref{kapexp}) and (\ref{chiexp}). The leading order of this expansion is the configuration presented in the previous subsection. For the subleading orders it is more difficult to find results exact in $m$. We present the expansions in \ref{cnonulo}.

In the expansions the large $m$ limit corresponds to large $\kappa$, which is enough to establish a comparison with perturbative weak coupling results. Moreover, It turns out that large $m$ and small $c$ implies small $\chi$, thus including the case $\chi=0$  which particularly interesting because it preserves supersymmetry.
We find
\begin{align}
m =& \left(\frac{R\kappa}{L}-\frac{L}{2R\kappa}-\frac{L(16L^2+4\pi^2R^2+5\pi^2 L^2)}{128 R^3 \kappa^3}+ {\cal O}(\kappa^{-5})\right)\label{mexp}
\\
&+\chi \left(\frac{\pi(2R^2+3L^2)}{8LR\kappa}+\frac{\pi L (92 R^2+107L^2)}{128 R^3\kappa^3}+ {\cal O}(\kappa^{-5})\right)
-\chi^2 \left(\frac{R\kappa}{2L}+ \frac{6R^2+7L^2}{4LR\kappa}+{\cal O}(\kappa^{-3})\right)\,,\nonumber
\\
c=&-\left(\frac{\pi L}{8 R^2\kappa^2}+\frac{15\pi L^3}{128 R^4\kappa^4} + {\cal O}(\kappa^{-6})\right)
+\chi \left(\frac{1}{L}+\frac{L}{2R^2\kappa^2} +\frac{3L(4L^2+\pi^2R^2+2\pi^6L^2)}{32R^4\kappa^4} + {\cal O}(\kappa^{-6})\right)\nonumber\\
&-\chi^2 \left(\frac{\pi(R^2+4L^2)}{8LR^2\kappa^2}+\frac{3\pi L(19R^2+34L^2)}{64R^4\kappa^4}+ {\cal O}(\kappa^{-6})\right)
+{\cal O}(\chi^3)\,,
\label{cexp}
\end{align}

Using the expansions in the regularized on-shell action we obtain
\begin{align}
\label{action2ndorder}
S = &
-\frac{\pi R k \chi }{L} -
\frac{\lambda L}{8 R k} \left[1-\frac{4\chi}{\pi}+\chi^2\left(\frac{R^2}{L^2}+\frac{5}{2}\right)+\mathcal{O}(\chi^3)\right]
\\
&-
\frac{\lambda ^2 L^3}{128 \pi ^2 k^3 R^3}
\left[
5- \frac{4\chi}{\pi} \left (4+\pi^2\left(\frac {R^2} {L^2} + \frac{7}{4} \right) \right) +
 \chi^2\left(\frac{94 R^2}{L^2} + \frac {233}{2}\right) +\mathcal{O}(\chi^3)
\right]
+k \mathcal{O}\left(\frac{\lambda^3}{k^6}\right)\nonumber
\end{align}
In (\ref{action2ndorder}) we have expanded up to $\lambda^2$ and up to $\chi^2$. To go beyond the order in $\chi^2$ one would need to solve beyond the order $c^2$. The first line in (\ref{action2ndorder}) will be contrasted with the 1-loop perturbative computation.

\section{Perturbative computation}
\label{perturbative}

Now we focus our attention to the gauge theory in order to compute the Wilson loop in perturbation theory. Some of the results in this section are similar to the ones obtained in \cite{KristWL} for the straight line case.

The interface at $x_3=0$ connects two gauge theories with gauge groups $SU(N)$ (say $x_3<0$) and $SU(N-k)$ ($x_3>0$). This is achieved by letting 3 scalar fields of ${\cal N}=4$ SYM, which we will take to be  $\Phi_1$, $\Phi_2$ and
$\Phi_3$, acquire non-trivial expectation values at the classical level
for $k$ of their components. To do this in a supersymmetric fashion, the classical vacuum expectation values are given by the fuzzy funnel solution \cite{Constable:1999ac} to the Nahm's equations \cite{Nahm:1977tg}.
\be
\langle \Phi_i\rangle_{cl} = -\frac{1}{x_3}t_i \oplus 0_{(N-k)\times (N-k)}\,, \qquad i=1,2,3
\label{fuzzy}
\ee
where the $\{t_i\}$ form a $k$-dimensional representation of the $\text{SU}(2)$ algebra (see Appendix \ref{alg} and \cite{Constable:1999ac,Nagasaki}).
Consequently, mass-like terms for some components of the quantum fields arise after expanding the action around the classical value of the fields. The diagonalization of the color structure of the quadratic terms that provides the mass spectrum was worked out in \cite{Kristonepoint,Buhl-Mortensen:2016jqo}  (for completeness we present  the data  in Appendix \ref{prop}). The resulting equation for the scalar propagator is of the form \footnote{Since fermionic modes do not contribute in 1-loop computation, we do not present
the corresponding propagators.}
\be
\left(-\partial^\mu\partial_{\mu}+\frac{m^2}{(x_3)^2}\right)K(x,y)=
\frac{g_{YM}^2}{2}\delta(x-y)\,,
\ee
where  $\frac{m}{x_3}$ is the mass for each scalar mode, coming from the VEV of the scalars of the fuzzy funnel solution, $m$ should not be confused with the parameter $m$ in the gravity computation, the values of $m$ for each scalar mode are reported in in Appendix \ref{prop}. The above propagator can be solved in terms of the AdS propagator (see Appendix \ref{prop}). We are concerned with the 1-loop correction for the the expectation value of the circular Wilson loop \eqref{wl}, which is at the distance $L$ from the defect and has a radius $R$. Because of presence of $\Phi_3$, the exponent has a non-trivial classical value. Expanding around it and keeping terms up to 1-loop order we obtain
\begin{align}
\langle W\rangle = & \, \langle W\rangle_{(0)} + \langle W\rangle_{(1)} + \langle W\rangle_{(2)}\nonumber
\\
= & \, \text{tr}U^{cl}(0,2\pi)  + R\int_0^{2\pi} d\alpha\langle\text{tr}U^{cl}(0,\alpha)\mathcal{A}(\alpha)U^{cl}(\alpha,2\pi)\rangle
\nonumber
\\
&  + R^2\int_0^{2\pi} \!d\alpha\int_\alpha^{2\pi}\!d\beta
\langle\text{tr}U^{cl}(0,\alpha)\mathcal{A}(\alpha)U^{cl}(\alpha,\beta)\mathcal{A}(\beta)U^{cl}(\beta,2\pi)\rangle\,,
\label{wlexp}
\end{align}
where
\be
U^{cl}(\alpha,\beta)=\exp\left(-R\sin\chi\int_\alpha^\beta d\tau \langle\Phi_3\rangle_{cl}\right)=
\exp\left(\frac{(\beta-\alpha)R\sin\chi}{L}t_3\right)\,.
\label{cl}
\ee

For the classical contribution $\langle W\rangle_{(0)}$ we have to perform the trace of \eqref{cl} with $\alpha=0$ and $\beta=2\pi$ (for conventions on the algebra generators we refer to the  Appendix \ref{alg}). In particular we can see that $E^i_iE^j_j=\delta{ij}E^i_i$ and $\text{tr}E^i_j=\delta_{ij}$, therefore
\be
\langle W\rangle_{(0)} = (N-k)+
\sum_{l=1}^k e^{\frac{2\pi R\sin\chi}{L}d_{k,l}}=(N-k)+\frac{\sinh\left(\frac{\pi R\sin\chi}{L}k\right)}{\sinh\left(\frac{\pi R\sin\chi}{L}\right)}\,.
\label{tree}
\ee

The second term in \eqref{wlexp}, which we refer to as $\langle W\rangle_{(1)}$, reads
\be
\langle W\rangle_{(1)}=R\int_0^{2\pi}d\alpha\left(e^{\frac{\alpha R\sin\chi}{L}t_3}\right)_{ab}\langle \mathcal{A}(\alpha)\rangle_{bc}^{\text{1-loop}}
\left(e^{\frac{(2\pi-\alpha)R\sin\chi}{L}t_3}\right)_{ca},
\label{lol}
\ee
where indices $a,b,c$ run from 1 to $k$ and summation over repeated indices is implied. The 1-point function at 1-loop has already been computed \cite{Kristonepoint} finding that it vanishes after regularization
\be
\langle\mathcal{A}(\alpha)\rangle^{\text{1-loop}}=0\,.
\label{onepoint}
\ee
Therefore, $\langle W\rangle_{(1)}$ is trivially vanishing.

The last contribution to \eqref{wlexp} is $\langle W\rangle_{(2)}$. We decompose this contribution using the mass spectrum structure presented in table \ref{mass} in Appendix \ref{prop}
\be
\langle W\rangle_{(2)} = T_1 + T_2 + T_3 + T_4\,,
\ee
where
\begin{align}
T_1 &= R^2  \int_0^{2\pi} d\alpha\int_\alpha^{2\pi}d\beta\left\langle\left(e^{\frac{\alpha\sin\chi}{L}t_3}\right)_{ab}\mathcal{A}_{bc}(\alpha) \left(e^{\frac{(\beta-\alpha)\sin\chi}{L}t_3}\right)_{cd}\mathcal{A}_{de}(\beta)\left(e^{\frac{(2\pi-\beta)\sin\chi}{L}t_3}\right)_{ea}\right\rangle\,,\\
T_2 &= R^2  \int_0^{2\pi} d\alpha\int_\alpha^{2\pi}d\beta\left\langle\left(e^{\frac{\alpha\sin\chi}{L}t_3}\right)_{ab}\mathcal{A}_{bi}(\alpha) \mathcal{A}_{ic}(\beta)\left(e^{\frac{(2\pi-\beta)\sin\chi}{L}t_3}\right)_{da}\right\rangle\,,\\
T_3 &= R^2  \int_0^{2\pi} d\alpha\int_\alpha^{2\pi}d\beta\left\langle\mathcal{A}_{ia}(\alpha) \left(e^{\frac{(\beta-\alpha)\sin\chi}{L}t_3}\right)_{ab}\mathcal{A}_{bi}(\beta)\right\rangle\,,\\
T_4 &= R^2 \int_0^{2\pi} d\alpha\int_\alpha^{2\pi}d\beta\left\langle\mathcal{A}_{ij}(\alpha)\mathcal{A}_{ji}(\beta)\right\rangle\,,
\end{align}
where $a,b,c,d,e=1,\ldots,k$ and $i,j=k+1,\ldots, N$.

$T_1$ involves only matrix elements of the $(k+1)\times (k-1)$ block of the color matrices. The total number of modes amounts to the dimension of the adjoint representation of $\text{SU}(k)$. Then this term is at most of order $k^2$, and therefore subleading in comparison with the others in the large $N$ limit.
On the other hand, $T_4$ amounts to the contribution of the non-massive modes, which lead to the well known $\mathcal{N}=4$ computation but now with $N$ replaced by $(N-k)$.
{From the dual string theory point of view these terms should come from string solutions that do not end on the D5-brane. 
Since we are interested in $\frac{\lambda}{k^2}$ dependent corrections, we will not focus on this contribution.  }

The arguments just exposed leave $T_2$ and $T_3$ as the possible sources of $\frac{\lambda}{k^2}$ corrections. Thus, we will focus on them in order to compare with our classical string theory results presented in section \ref{string}. They involve the non-diagonal block terms. 
We make use of the propagators and the  $k$-dependent mass spectrum in Appendix \ref{prop} and \eqref{mass}. Using also the form of the $t_3$ generator we find
\begin{align}
T_2 &=R^2\int_0^{2\pi} d\alpha\int_\alpha^{2\pi}d\beta \sum_{a=1}^k e^{\frac{2\pi-\beta+\alpha}{L}R\sin\chi d_{k,a}}\left\langle\mathcal{A}_{ai}(\alpha)\mathcal{A}_{ia}(\beta)\right\rangle\,,
\label{t2}\\
T_3 &=R^2\int_0^{2\pi} d\alpha\int_\alpha^{2\pi}d\beta \sum_{a=1}^k e^{\frac{\beta-\alpha}{L}R\sin\chi d_{k,a}}\left\langle\mathcal{A}_{ia}(\alpha)\mathcal{A}_{ai}(\beta)\right\rangle\,.
\label{t3}
\end{align}
Using the fields in the diagonal basis and the mass spectrum \eqref{mass}, the corresponding expectation value results in
\begin{align}
\left\langle\mathcal{A}_{ai}(\alpha)\mathcal{A}_{ib}(\beta)\right\rangle = \left\langle\mathcal{A}_{ia}(\alpha)\mathcal{A}_{bi}(\beta)\right\rangle =&\, \delta_{ab}(N-k)(1-\cos(\beta-\alpha))K_{\frac{k}{2}}(\alpha,\beta)
\label{vev}
\\
 +& \delta_{ab}\frac{(N-k)}{2k}\sin^2\chi
 \left((k-1)K_{\frac{k+2}{2}}(\alpha,\beta)+(k+1)K_{\frac{k-2}{2}}(\alpha,\beta)\right)\,,
\nonumber
\end{align}
where $K_{\nu}(\alpha,\beta)$ is the propagator defined in \eqref{massprop}. We can compute the angular integral in \eqref{massprop} using $\lvert\vec{x}(\alpha)-\vec{x}(\beta)\rvert=2R\sin\frac{\beta-\alpha}{2}$ and defining $r=\lvert\vec{k}\rvert$
\be
K_{\nu}(\alpha,\beta)=\frac{g^2_{YM}L}{8\pi^2R}\int\limits_0^\infty dr r \frac{\sin\left(2Rr\sin\frac{\beta-\alpha}{2}\right)}{\sin\frac{\beta-\alpha}{2}}
I_\nu(rL)K_\nu(rL)\,.
\ee
It is not difficult to do the sums over $a$ and one of the angular integrals because the integrands in \eqref{t2} and \eqref{t3} depend on $\alpha$ and $\beta$ through the difference. Collecting both contributions we find
\be
T_2+T_3=(N-k)\tfrac{g^2_{YM}R}{4\pi L}\int\limits_0^\infty dr r\int\limits_0^{\pi}d\delta\left(\frac{\sinh\left(\frac{(\pi-\delta)R\sin\chi}{2L}k\right)}{\sinh\left(\frac{(\pi-\delta)R\sin\chi}{2L}\right)}
+\frac{\sinh\left(\frac{(\pi+\delta)R\sin\chi}{2L}k\right)}{\sinh\left(\frac{(\pi+\delta)R\sin\chi}{2L}\right)}\right)
\left(\mathcal{I}_1+\sin^2\chi\mathcal{I}_2\right)
\label{int}
\ee
where
\begin{align}
\mathcal{I}_1 &=2\cos\frac{\delta}{2}\sin\left(\frac{2Rr}{L}\cos\frac{\delta}{2}\right)
I_{\frac{k}{2}}(r)K_{\frac{k}{2}}(r)\,,
\\
\mathcal{I}_2 &=\frac{\sin\left(\frac{2Rr}{L}\cos\frac{\delta}{2}\right)}{\cos\frac{\delta}{2}}
\left(\frac{k-1}{2k}I_{\frac{k+2}{2}}(r)K_{\frac{k+2}{2}}(r)+\frac{k+1}{2k}
I_{\frac{k-2}{2}}(r)K_{\frac{k-2}{2}}(r)-I_{\frac{k}{2}}(r)K_{\frac{k}{2}}(r)\right)\,,
\end{align}
where we have rescaled the $L$ dependence from the Bessel functions.

The integrals involved in \eqref{int} are difficult to solve analytically.
In the limit $L/R\to 0$ the problem remains non-trivial but it becomes simpler.
In that limit we have
\be
\begin{array}{rll}
\frac{\sinh\left(\frac{\pi R\sin\chi}{L}k\right)}{\sinh\left(\frac{\pi R\sin\chi}{L}\right)}&\sim & e^{ 2\pi\eta}\,,\\
\left(\frac{\sinh\left(\frac{(\pi-\delta)R\sin\chi}{2L}k\right)}{\sinh\left(\frac{(\pi-\delta)R\sin\chi}{2L}\right)}+\frac{\sinh\left(\frac{(\pi+\delta)R\sin\chi}{2L}k\right)}{\sinh\left(\frac{(\pi+\delta)R\sin\chi}{2L}\right)}\right)&\sim & e^{\left(\pi+\delta\right)\eta}\,,
\end{array}
\ee
where we have conveniently defined $\eta=\frac{R\sin\chi}{2L}(k-1)$. Then, in this limit, \eqref{int} reduces to
\be
T_2+T_3 \sim (N-k)\frac{g^2_{YM}R}{4\pi L} e^{\pi\eta}\int\limits_0^\infty dr r\int\limits_0^{\pi}d\delta e^{\eta\delta}
\left(\mathcal{I}_1+\sin^2\chi\mathcal{I}_2\right)\,.
\label{x3int}
\ee
Using an identity of Bessel functions presented in appendix \ref{I2} one can integrate by parts the $r$ integral of $\mathcal{I}_2$ and get
\begin{align}
T_2+T_3  & \sim (N-k)\tfrac{g^2_{YM}R}{2\pi L} e^{\pi\eta}
\int\limits_0^\infty dr r I_{\frac{k}{2}}(r)K_{\frac{k}{2}}(r)\int\limits_0^{\pi}d\delta e^{\eta\delta}
\cos\tfrac{\delta}{2}\sin\left(\tfrac{2Rr}{L}\cos\tfrac{\delta}{2}\right)
\\
&- (N-k)\tfrac{g^2_{YM}R}{2\pi L} \sin^2\chi
e^{\pi\eta}\int\limits_0^\infty dr \left(\tfrac{1}{2}-rI'_{\frac{k}{2}}(r)K_{\frac{k}{2}}(r)-\tfrac{1}{2} I_{\frac{k}{2}}(r)K_{\frac{k}{2}}(r)\right)
\int\limits_0^{\pi}d\delta e^{\eta\delta}
\cos\left(\tfrac{2Rr}{L}\cos\tfrac{\delta}{2}\right)\nonumber
\end{align}
Now we have to compute the $\delta$-integrals which is is also done in the appendix \ref{I2}. In the large $\eta$ limit one can see that
\begin{align}
\int\limits_0^\pi d\delta e^{\eta\delta}\cos\left(\tfrac{2Rr}{L}\cos\frac{\delta}{2}\right)
& \sim \frac{\eta e^{\pi\eta}}{\left(\tfrac{Rr}{L}\right)^2+\eta^2}\,,
\label{cosint}
\\
\int\limits_0^\pi d\delta e^{\eta\delta}\cos\tfrac{\delta}{2}\sin\left(\tfrac{2Rr}{L}\cos\frac{\delta}{2}\right)
& \sim \frac{ \left(\frac{Rr}{L}\right)\eta e^{\pi\eta}}{\left(\left(\tfrac{Rr}{L}\right)^2+\eta^2\right)^2}\,.
\label{sinint}
\end{align}

Therefore in this limit we obtain
\begin{align}
\begin{split}
T_2+T_3 \sim \tfrac{\lambda R}{2\pi L}e^{2\pi\eta}&\left[\left( \tfrac{L}{R}\right)^3\int\limits_0^{\infty}dr\frac{\eta r^2}{\left(r^2+\left(\tfrac{\eta L}{R}\right)^2\right)^2}I_{\frac{k}{2}}(r)K_{\frac{k}{2}}(r)\right.\\
&-\left.\sin^2\chi\left(\tfrac{L}{R}\right)\int\limits_0^{\infty}dr\frac{\eta }{r^2+\left(\tfrac{\eta L}{R}\right)^2}
\left(\tfrac{1}{2}-rI'_{\frac{k}{2}}(r)K_{\frac{k}{2}}(r)-\tfrac{1}{2} I_{\frac{k}{2}}(r)K_{\frac{k}{2}}(r)\right)\right]
\end{split}
\label{leading}
\end{align}
where we have taken the large $N$ limit and introduced the 't Hooft coupling $\lambda=g_{YM}^2 N$. Note that the second line leads to the result obtained in \cite{KristWL} but making the replacement $T\to 2\pi$.

Rescaling the integration variable to $u=\frac{2r}{k}$ and expanding for large $k$, the first term in \eqref{leading} becomes
\be
\frac{\lambda}{\pi k}\left(\frac{L}{R}\right)^2e^{ 2\pi\eta}\int\limits_0^\infty dr\frac{\eta u^2}{\left((u^2+(\tfrac{2\eta L}{R k})^2\right)^2\sqrt{1+u^2}}
= \frac{\lambda L}{4\pi R k }\frac{e^{\frac{(k-1)\pi R}{L}\sin\chi}}{\cos^3\chi}\left(\frac{\pi}{2}-\chi-\frac12 \sin 2\chi\right)\,,
\label{w2a}
\ee
where we have replaced $\eta=\frac{\sin\chi(k-1)R}{2L}$. The remaining term in \eqref{leading}, expanded for large $k$, is
\be
\frac{\lambda R}{4\pi L k^2}\eta e^{2\pi\eta}\int\limits_0^\infty\frac{du}{\left(u^2+(\tfrac{2\eta L}{R k})^2\right)\left(1+u^2\right)^{\frac32}}=
\frac{\lambda R}{4\pi L k} e^{\frac{(k-1)\pi R}{L}\sin\chi}\frac{\sin^2\chi}{\cos^3\chi}\left(\frac{\pi}{2}-\chi-\frac12 \sin 2\chi\right)
\label{w2b}
\ee
We are now in a position to collect all the contributions to $\langle W\rangle$. At this point it is instructive to distinguish between different sorts of contributions. At tree level, already for for large $R/L$  and large $k$, we can define
\be
\langle W\rangle _{(0)}^{I} = N-k\,,\qquad
\langle W\rangle _{(0)}^{II} = e^{\frac{(k-1)\pi R}{L}\sin\chi}\,.
\ee
Accordingly, at 1-loop order we can define
\be
\langle W\rangle _{(2)}^{I} = T_4\,,\qquad
\langle W\rangle _{(2)}^{II} = T_2+T_3\,.
\ee
By comparison with semiclassical computations we realize that contributions $\langle W\rangle ^{I}$
and $\langle W\rangle^{II}$ correspond to different saddle point approximations of the string theory partition functions. More precisely, $\langle W\rangle ^{I}$ accounts for the usual configuration in which the string does not end on the D5-brane, while $\langle W\rangle ^{II}$ accounts for the configuration found in section \ref{string} in which the string do end on the D5-brane
\footnote{ Calling $S_I$ and $S_{II}$ the corresponding on-shell actions
$$
\langle W\rangle^I + \langle W\rangle^{II} \simeq (N-k)e^{S_I}+e^{S_{II}}
$$
The term $e^{S_I}$ comes from a string extending between a D3-brane and a stack of $(N-k)$ D3-branes, which
would explain the weighting factor $(N-k)$.}. Collecting the contributions from (\ref{tree}), (\ref{w2a}) and (\ref{w2b}) we then have, for large $R/L$  and large $k$,
\be
\log \langle W\rangle ^{II}\simeq \frac{k\pi R}{L}\left(\sin\chi+\frac{\lambda}{4\pi^2 k^2}\frac{1}{\cos^3\chi}\left(\frac{\pi}{2}-\chi-\frac12 \sin 2\chi\right)\left(\sin^2\chi+ \left(\frac{L}{R}\right)^2\right)\right)\,.
\label{pert}
\ee
In order to compare with the strong coupling result presented in section \ref{string}, we expand (\ref{pert}) for small $\chi$ thus obtaining
\be
\log \langle W\rangle ^{II}
\simeq\frac{\pi R k}{L}\left[\chi+\frac{\lambda}{8\pi}\left(\frac{L}{R k}\right)^2\left(1-\frac{4\chi}{\pi}+\chi^2\left(\frac{R^2}{L^2}+\frac{3}{2}\right)\right)\right]\,.
\ee
This is in agreement with \eqref{action2ndorder}. The only apparent difference is the $\frac{3}{2}$ in the term order $\chi^2$. However, this is a subleading contribution in the large $R/L$ expansion and as such is out of the range of validity of the perturbative computation. A further computation of the subleading corrections of the Feynman diagram should reproduce the full $\chi^2$ term coming from the string theory computation.

\section{Discussion}

We have studied circular Wilson loops in presence of a codimension one defect that acts as an interface between two gauge theories with $\text{SU}(N)$ and $\text{SU}(N-k)$ gauge groups respectively. We computed both 1-loop perturbative expectation values in gauge theory and the corresponding semiclassical string theory partition functions.  Quite interestingly, in this example we identified different hierarchies for different types of contributions to $\langle W\rangle$, which should be associated to different semiclassical saddle points of the string theory partition function.

At the end, we have performed the double scaling limit proposed in \cite{Nagasaki} and concluded that in this case one can also extrapolate weak coupling results to the strong coupling limit. We have checked the extrapolation of the 1-loop results. Moreover the on-shell action in section \ref{string} was computed
up to order $\left(\lambda/k\right)^2$ in eq. (\ref{action2ndorder}), thus providing a prediction for
$\log\langle W\rangle^{II}$ at 2-loop order.

We have also considered whether the circular Wilson loop is supersymmetric or not. We relegated the details to the appendix \ref{susy} and have found that for $\chi =0$ the operator preserves half of the supersymmetries of the defect conformal field theory. From the string theory point of view $\chi(c,m)=0$ corresponds to a specific relation between parameters $c$ and $m$. We could systematically obtain an order by order expansion
for this supersymmetric configuration but it would be very useful to find it exactly thus obtaining an all loop order prediction for its expectation value.

Alternatively, one might wonder whether the supersymmetric Wilson loop can be exactly computed using localization techniques, which would provide an ideal scenario for a precision test, as it was the case for this kind of circular Wilson loops in $\mathcal{N}=4$ SYM \cite{Pestun:2007rz}. One can proceed first by mapping the supercharges of flat space, the spinor solutions preserving the Wilson loop and the defect parametrized by  \eqref{eq:Def+WLsusy}, to the sphere  as in \cite{Pestun:2007rz}. Conformal invariance requires an additional coupling for the scalars with scalar curvature. The defect, being half-BPS, can be placed in an  S$^3$ at the equator of S$^4$. Conformal invariance requires that fundamental scalar fields living on the defect has to be coupled to the S$^3$ scalar curvature as well. The action for the defect conformal field theory that was worked out in \cite{DeWolfe:2001pq} has to be generalized to account for the case $k\neq0$.
Because of the flux through the S$^2$ factor of the dual D5-brane solutions, the radius of its AdS$_4$ factor  will be different to the radius of and AdS$_5$
and dependent on $k$ \cite{Karch:2000gx}. Therefore, the action of the defect will bring in effective couplings depending on the flux along $\text{S}^2$ (on the classical fuzzy funnel solution). An important aspect of the computation in \cite{Pestun:2007rz}  is related to the non-perturbative contributions that  come from instantons and anti-instantons localized at the poles of the $\text{S}^4$. The theory living in the defect couples to the gauge multiplet, that will bring additional features in contrast with the theory without defect. One expects non-perturbative contributions of SU$(N)$ from one side of the defect and SU$(N-k)$ from the other, therefore the non-perturbative contributions have to be understood.

\section*{Acknowledgements}
V. I. Giraldo-Rivera is  very grateful with the members of IFLP  where this project started and for their hospitality. We would like to thank Diego Trancanelli for helpful comments. D.H.Correa and J.Aguilera-Damia  are supported by CONICET and grants PICT 2012-0417 and PIP 0681.

\appendix

\section{String configuration with $c\neq 0$}
\label{cnonulo}
For generic values of $m$ and $c$, the equations of motions for string configurations are
\be
y y'' + 2\left(r'\right)^2 + \frac{m^2}{1+m^2} y^2 + 2(x_3')^2=0\,,\quad
y r'' -2 y' r' - \frac{yr}{1+m^2}=0\,,\quad x_3' + c y^2 =0\,,\quad \theta'=m\,,
\ee
subject to the following boundary conditions at the boundary
\be
y(0)=0\,,\qquad r(0)=R\,,\qquad x_3(0)=L\,,\qquad \theta(0)=\chi\,,
\label{bc bound}
\ee
and the maximum value for the variable $\sigma$
\be
y'(\tilde{\sigma})-\kappa cy^2(\tilde{\sigma})=0\,,\qquad
 r' (\tilde{\sigma}) =0\,,\qquad
y(\tilde{\sigma})-\frac{1}{\kappa}x_3(\tilde{\sigma})=0\,,\qquad
\theta(\tilde{\sigma})=\frac{\pi}{2}\,.
\label{bc brane}
\ee

This is a complicated system of non-linear differential equations. However, since we know the solution for $c=0$, we can expand the general solution in powers of $c$
\be
\begin{array}{rcl}
y(x)&=&y_0(x)+c y_1(x)+c^2 y_2(x) + {\cal O}(c^3)\,,\\
r(x)&=&r_0(x)+c r_1(x)+c^2 r_2(x) + {\cal O}(c^3)\,,\\
x_3(x)&=&x_{3,0}(x)+c x_{3,1}(x)+c^2 x_{3,2}(x) + {\cal O}(c^3)\,,
\label{expansionapp}
\end{array}
\ee
which leads to a system of linear differential equations. Solving them and imposing the boundary conditions order by order we obtain
\begin{align}
y_0(x)&=\frac{R}{\sqrt{1+m^2}} \;\text{sn}\left(x,\tfrac{1}{1+m^2}\right)\,,\nonumber
\\
r_0(x)&= R \;\text{dn} \left(x,\tfrac{1}{1+m^2}\right)\,,
\label{solution0}
\\
x_{3,0} &= L\,,\nonumber
\end{align}
and
\begin{align}
y_1(x)&= L y_0(x)\left[x- \mathbb{E}({\rm am}(x,\tfrac{1}{1+m^2}),\tfrac{1}{1+m^2}))\right]\,,\nonumber
\\
r_1(x)&= L r_0(x)\left[x- \mathbb{E}({\rm am}(x,\tfrac{1}{1+m^2}),\tfrac{1}{1+m^2}))\right]\,,
\label{solution1}
\\
x_{3,1}(x)&= -R^2 \left[x- \mathbb{E}({\rm am}(x,\tfrac{1}{1+m^2}),\tfrac{1}{1+m^2}))\right]\,.\nonumber
\end{align}
For leading and next to leading order these solutions are exact in $m$. For the next to next to leading order, equations are more complicated and we have solved them expanding for large $m$,
\begin{align}
y_2(x)&= \frac{1}{m^3}y_2^{(3)}(x)+\frac{1}{m^5}y_2^{(5)}(x)+{\cal O}(m^{-7})\,,
\nonumber
\\
r_2(x)&= \frac{1}{m^4}r_2^{(4)}(x)+\frac{1}{m^6}r_2^{(6)}(x)+{\cal O}(m^{-8})\,,
\label{solution3}
\\
x_{3,2}(x)&= \frac{1}{m^4}x_{3,2}^{(4)}(x)+\frac{1}{m^6}x_{3,2}^{(6)}(x)+{\cal O}(m^{-8})\,,\nonumber
\end{align}
and we have found
\begin{align}
y_2^{(3)}(x) =& -\frac{R}{16} \left(R^2+L^2\right) (9 \sin x+\sin{3 x}-12 x \cos x)\,,
\nonumber
\\
y_2^{(5)}(x) =&
\frac{R}{128}\left[ (95 R^2 + 99L^2) \sin x + 8 x^2 (R^2 + 5L^2)\sin x
-(R^2+3 L^2) \sin 5x\right.\nonumber
\\
& \qquad \left.- 4x(25 R^2+29 L^2) \cos x + 16 x (R^2+2L^2) \cos 3x
-2 R^2 \sin 3x \right]\,,
\nonumber
\\
r_2^{(4)}(x) =& -\frac{R}{64}
\left[8x^2(R^2-L^2)-17R^2-19L^2
+16 x \left(R^2+2 L^2\right) \sin{2x}\right.\nonumber
\\
&\qquad\quad\left.+16 \left(R^2+L^2\right)
\cos{2x}+\left(R^2+3L^2\right) \cos{4x}\right]\,,
\nonumber
\\
r_2^{(6)}(x) =&\frac{R}{2048}
\left[-760 R^2-872 L^2-x(382R^2+498L^2)+512x^2(R^2-L^2)
-80 x (R^2 + 3 L^2) \sin 4x
\right.
\nonumber
\\
&+
4(193 R^2 + 197 L^2)\cos 2x +192 x^2 (R^2 + 3 L^2) \cos 2x
-8 (R^2 - 13 L^2) \cos 4x
\nonumber
\\
&\left.
- 4 (R^2 + 5 L^2) \cos 6x + (191 R^2 + 249 L^2) \sin 2x + 4 x (3 R^2 + 118 L^2) \sin 2x
\right]\,,
\nonumber
\\
x_{3,2}^{(4)}(x) =& - \frac{R^2L}{16}\left(2x-\sin 2x \right)^2\,,
\nonumber
\\
x_{3,2}^{(6)}(x) =& \frac{R^2L}{64}\left((6-\cos 2x)\sin^2 2x-x(26\sin 2x-3\sin 4x+4x^2(7-2\cos 2x)) \right)\,.
\end{align}

And from the boundary conditions we have
\begin{align}
\tilde x =& \mathbb{K}\left(\tfrac{1}{1+m^2}\right)- c L -
c^2\left[\frac{\pi(R^2+3L^2)}{8m^2}-\frac{\pi(R^2-9L^2)}{64m^4}+{\cal O}(m^{-6})\right]+{\cal O}(c^3)\,,\nonumber
\\
\frac{\pi k}{\sqrt{\lambda}} =& \frac{L}{R}\sqrt{1+m^2} + c \frac{R^2+L^2}{R}\sqrt{1+m^2}
\left[\mathbb{E}\left(\tfrac{1}{1+m^2}\right)-\mathbb{K}\left(\tfrac{1}{1+m^2}\right)\right]\nonumber
\\
 &+c^2\left[\frac{L^3 m}{2R} + \frac{L}{4m R}(6 R^2 + 5 L^2) +{\cal O}(m^{-3})\right]
 +{\cal O}(c^3)
\,,
\end{align}

\section{$k$-dimensional SU(2) generators}\label{alg}

Let $E^i_j$ be $k\times k$ matrices such that
\be
E^i_j E^k_l = \delta^k_j E^i_l\,.
\ee
We can represent them by taking
\be
(E^i_j)_{ab}= \delta_{ia}\delta_{jb}\,.
\ee
In terms of these matrices, we can represent the $\text{SU}(2)$ algebra as
\be
t_{+}=\sum_{i=1}^{k-1}c_{k,i}E^i_{i+1}\,,\qquad
t_{-}=\sum_{i=1}^{k-1}c_{k,i}E^{i+1}_i\,,\qquad
t_3=\sum_{i=1}^{k}d_{k,i}E^i_i\,,
\ee
with
\be
c_{k,i}=\sqrt{i(k-i)}\,,\qquad d_{k,i}=\frac12(k-2i+1)\,.
\ee

\section{Massive proagators}\label{prop}

In terms of the AdS$_4$ propagator, satisfying
\be
\left(-\nabla^\mu\nabla_\mu+\tilde{m}^2\right)K_{AdS}(x,y)=\frac{\delta(x-y)}{\sqrt{g}}\,,
\ee
one can define
\be
K(x,y)=\frac{g_{YM}^2K_{AdS}(x,y)}{2x_3y_3}\,,
\ee
which is a solution for
\be
\left(-\partial^\mu\partial_{\mu}+\frac{m^2}{(x_3)^2}\right)K(x,y)
=\frac{g_{YM}^2}{2}\delta(x-y)\,,
\ee
provided $\tilde{m}^2=m^2-2$. We use the following integral representation for the propagator
\be
K_\nu(x,y)=\frac{g^2_{YM}\sqrt{x_3y_3}}{2}
\int\frac{d^3 \vec{k}}{(2\pi)^3} e^{-i \vec{k}\cdot(\vec{x}-\vec{y})}
I_\nu(\lvert\vec{k}\rvert x_3)K_\nu(\lvert\vec{k}\rvert y_3)\,,
\label{massprop}
\ee
where $\vec{k}$, $\vec{x}$ and $\vec{y}$ are 3d vectors in the $(x_0,x_1,x_2)$ directions, $I_\nu$ and $K_\nu$ are Bessel functions and $\nu$ is related to the mass of the propagating mode
\be
\nu=\sqrt{m^2+\frac14}\,.
\ee

The diagonalization fo the mass matrix coming from the action by expanding the lagrangian around the classical solution was obtained in \cite{Kristonepoint,Buhl-Mortensen:2016jqo} . In the following table we report  the data that will be used in the main body of the paper.
\begin{table}[H]
\begin{center}
\ba
\begin{tabular}[H]{c|c|c|c}
$\text{Multiplicity}$ & $\nu\left(\Phi_{4,5,6},A_{0,1,2},c\right)$& $m \left(\psi_{1,2,3,4}\right)$ & $\nu\left(\Phi_{1,2,3},A_3,c\right)$\\ \hline
$j$& $j+\frac12$ & $j+1$ & $j+\frac32$ \\
$j+1$& $j+\frac12$ & $-j$ & $j-\frac12$ \\
$(k-1)(N-k)$& $\frac{k}{2}$ & $\frac{k+1}{2}$ & $\frac{k+2}{2}$ \\
$(k+1)(N-k)$& $\frac{k}{2}$ & $-\frac{k-1}{2}$ & $\frac{k-2}{2}$ \\
$(N-k)(N-k)$&$\frac12$&$0$&$\frac12$
\label{mass}
\end{tabular}
\ea
\end{center}
\caption{Mass Spectrum}
\end{table}
where $j=1,\ldots,k-1$.

\section{Some details for the perturbative computation}\label{I2}

By using of the following properties of Bessel functions
\be
\begin{array}{l}
I_{\nu\pm1}(z)=I'_{\nu}(z) \mp\left(\frac{\nu}{z}\right)I_{\nu}(z)\,,
\\
K_{\nu\pm1}(z)=-K'_{\nu}(z) \pm\left(\frac{\nu}{z}\right)K_{\nu}(z)\,,
\end{array}
\ee
we can relate the combination of Bessel functions appearing in the definition of $\mathcal{I}_2$ to a total derivative,
\be
z\!\left(\!I_{\nu}(z)K_{\nu}(z)-\tfrac{\nu-\tfrac{1}{2}}{2\nu}I_{\nu+1}(z)K_{\nu+1}(z)
-\tfrac{\nu+\tfrac{1}{2}}{2\nu}I_{\nu-1}(z)K_{\nu-1}(z)\!\right)
=\left(z I'_{\nu}(z)K_{\nu}(z)+\tfrac{1}{2}I_{\nu}(z)K_{\nu}(z)\right)'
\label{bessel}\nonumber
\ee
The integral of $\mathcal{I}_2$, in the large $\eta$ limit is proportional to
\begin{align}
\int\limits_0^\pi d\delta e^{\eta\delta}\cos\left(\tfrac{2Rr}{L}\cos\tfrac{\delta}{2}\right)  &=
\sum_{n=0}^{\infty}\int_0^\pi d\delta\frac{(-1)^n\left(\frac{2Rr}{L}\right)^{2n}}{(2n)!} e^{\eta\delta}\cos^n\tfrac{\delta}{2},\nonumber
\\
&\sim \sum_{n=0}^{\infty} \frac{(-1)^n\left(\frac{Rr}{L}\right)^{2n}e^{\eta\pi}}{\eta^{2n+1}} =
\frac{\eta e^{\eta\pi}}{\left(\frac{Rr}{L}\right)^2+\eta^2}\,.
\label{cosid}
\end{align}
On the other hand, the integral of $\mathcal{I}_1$, in the large $\eta$ limit is proportional to
\be
\int\limits_0^\pi d\delta e^{\eta\delta}\cos\tfrac{\delta}{2}\sin\left(\tfrac{2Rr}{L}\cos\tfrac{\delta}{2}\right)
\sim \frac{\frac{Rr}{L} \eta e^{\eta\pi}}{\left(\left(\frac{Rr}{L}\right)^2+\eta^2\right)^2}\,,
\ee
which is simply obtain by derivating (\ref{cosid}) with respect to $r$.

\section{Supersymmetry}\label{susy}

\subsection{Gauge theory}
In Euclidean signature the  most general   supersymmetric Wilson loop that has been considered until now is \cite{Zarembo:2002an,Drukker:1999zq}:
\begin{equation}
W=\tr \mathcal{P} \exp \big\{ \oint d\tau[ i A_\mu \dot{x}^\mu+\dot{y}^I\Phi_I]\big \},
\end{equation}
With $\dot{x}^2-\dot{y}^2=0$, and the constraint on the supersymmetry parameter.
\begin{equation}
\label{eq:WLsusy}
(i \Gamma^\mu \dot{x}_\mu +\rho^I \dot{y}_I)\epsilon(x)=0\,.
\end{equation}
The conventions used are those of the $\mathcal{N}=1$, 10d SYM dimensionally reduced, so $\Gamma$'s are  Dirac matrices of the 4d theory  and   $\rho$'s  act on the  $\text{SO}(6)_R$ indices of $\epsilon(x)$, $\Gamma$'s and $\rho$'s anti-commute. The general  spinor parameter is given by:
\begin{equation}
\epsilon(x)=\epsilon_0+x^\mu \Gamma_\mu \epsilon_1\,,
\end{equation}
where $\epsilon_0$ and $\epsilon_1$ are constant spinors  corresponding to  Poincare supercharges  and  special conformal supercharges respectively.\\
The Wilson loop we considered has the  following parametrization
\begin{eqnarray}
x^\mu(\tau)=(0,R \cos\tau,R \sin\tau,L)\quad\text{and}\quad \dot{y}^I=|\dot{x}|(0,0,-\sin\chi,0,0,-\cos\chi)\,,
\end{eqnarray}
then \eqref{eq:WLsusy} is
\begin{equation}
R(-i \Gamma^1 \sin\tau +i \Gamma^2 \cos\tau-\rho^3 \sin\chi -\rho^6 \cos\chi )\epsilon(x)=0\,,
\end{equation}
This has to be satisfied for all $\tau$ parametrizing the Wilson loop, so we have the following conditions
\begin{eqnarray}
&\sin\tau &:\quad\quad-i\Gamma^1\epsilon_0=[R(\sin\chi\rho^3+\cos\chi\rho^6)\Gamma^2+i L\Gamma^1\Gamma^3]\epsilon_1\,,\nonumber\\
&\cos\tau &:\quad\quad i\Gamma^2\epsilon_0=[R(\sin\chi\rho^3+\cos\chi\rho^6)\Gamma^1-i L \Gamma^2\Gamma^3] \epsilon_1\,,\nonumber\\
& 1& :\quad\quad(\sin\chi\rho^3+\cos\chi\rho^6)\epsilon_0=[-i R\Gamma^1\Gamma^2-L(\sin\chi\rho^3+\cos\chi\rho^6)\Gamma^3]\epsilon_1\,,\nonumber\\
&\sin\tau \cos\tau &:\quad\quad((\Gamma^2)^2-(\Gamma^1)^2)\epsilon_1=0\,, \nonumber\\
 &\cos^2\tau &:\quad\quad
(\Gamma^1\Gamma^2+\Gamma^1\Gamma^2)\epsilon_1=0\,.
\end{eqnarray}
The last two lines are trivially satisfied and these conditions are not all independent. Multiplying the first line by $\Gamma^2$ and the second by $\Gamma^1$, these two lines are shown to be the same
\begin{eqnarray}
&&i\Gamma^1\Gamma^2\epsilon_0=[-R(\sin\chi\rho^3+\cos\chi\rho^6)-i L \Gamma^1\Gamma^2\Gamma^3]\epsilon_1\,,\nonumber\\
&&(\sin\chi\rho^3+\cos\chi\rho^6)\epsilon_0=[-i R\Gamma^1\Gamma^2-L(\sin\chi\rho^3+\cos\chi\rho^6)\Gamma^3]\epsilon_1\,.
\end{eqnarray}
These las two equations are actually equivalent, either multiplying $-i\Gamma^2\Gamma^1$ by the first line  or $(\sin\chi\rho^3+\cos\chi\rho^6)$ by the second, we get
\begin{equation}
\label{eq:BPSWL}
\epsilon_0=-[i R(\sin\chi \rho^3+\cos\chi \rho^6)\Gamma^1\Gamma^2+L\Gamma^3]\epsilon_1\,.
\end{equation}
This means that this Wilson loop preserves half the number of supersymmetries, which are a mixed of Poincare and special conformal supercharges. We can write the final spinor  parameter for Wilson loop as
\ba
\label{eq:WLSCH}
\varepsilon ^{\text{WL}}(x^\mu(\tau))&=&\epsilon_0+ x^\mu(\tau)\Gamma_\mu\epsilon_1\,,
\nonumber\\
&=&-[i R(\sin\chi \rho^3+\cos\chi \rho^6)\Gamma^1\Gamma^2]\epsilon_1+ R \cos\tau \Gamma^1 \epsilon_1 +R \sin\tau\Gamma^2\epsilon_1\,.
\ea

We will follow \cite{CorreaYoung} where  the symmetries preserved by  $\mathcal{N}=4$ with the defect due to the presence of $\text{D}5$-brane were analized. To find the supersymmetries preserved  of the full system we have to further imposed the constraints \eqref{eq:BPSWL}. These constraints are given
by\footnote{In this notation $\Gamma^\mu=\gamma^\mu\otimes 1$ and $\rho^I=\gamma^5\otimes\gamma^I$ in \cite{CorreaYoung}.}
\begin{eqnarray}
&&P_+\epsilon_0=\epsilon_0 \quad\text{and}\quad P_+\epsilon_1=\epsilon_1\,,\nonumber\\
\text{with}&& P_+=\frac{1}{2}(1+\Gamma^3\rho^1\rho^2\rho^3)\,.
\end{eqnarray}
Notice that when $\chi=0$, the term coming from $\rho^3$ drops and the constraint is imposed by applying the projector on $\epsilon_1$. This is actually the only solution to the projector equation for the supersymmetry of the Wilson loop above (the projector equations comes with an overall $\sin\chi$). The full configuration is then 1/4 supersymmetric and is parametrized by the choices of $\epsilon_1$ which are also superconformal charges of the defect. The list of charges preserved by the defect can be written as
\begin{eqnarray}
&&(\uparrow,\uparrow\downarrow,\uparrow,\uparrow,\uparrow).Q\mp(\downarrow,\uparrow\downarrow,\downarrow,\downarrow,\downarrow).Q\,,
\nonumber\\
&&(\uparrow,\uparrow\downarrow,\uparrow,\uparrow,\downarrow).Q\pm (\downarrow,\uparrow\downarrow,\downarrow,\downarrow,\uparrow).Q\,,
\nonumber\\
&&(\uparrow,\uparrow\downarrow,\uparrow,\downarrow,\uparrow).Q\mp(\downarrow,\uparrow\downarrow,\downarrow,\uparrow,\downarrow).Q\,,
\nonumber\\
&&(\uparrow,\uparrow\downarrow,\uparrow,\downarrow,\downarrow).Q\pm(\downarrow,\uparrow\downarrow,\downarrow,\uparrow,\uparrow).Q\,,
\nonumber\\
&&(\downarrow,\uparrow\downarrow,\uparrow,\uparrow,\uparrow).S\mp(\uparrow,\uparrow\downarrow,\downarrow,\downarrow,\downarrow).S\,,
\nonumber\\
&&(\downarrow,\uparrow\downarrow,\uparrow,\uparrow,\downarrow).S\pm (\uparrow,\uparrow\downarrow,\downarrow,\downarrow,\uparrow).S\,,
\nonumber\\
&&(\downarrow,\uparrow\downarrow,\uparrow,\downarrow,\uparrow).S\mp(\uparrow,\uparrow\downarrow,\downarrow,\uparrow,\downarrow).S\,,
\nonumber\\
&&(\downarrow,\uparrow\downarrow,\uparrow,\downarrow,\downarrow).S\pm (\uparrow,\uparrow\downarrow,\downarrow,\uparrow,\uparrow).S\,.
\end{eqnarray}
The notation is as follows $(\uparrow\downarrow,\uparrow\downarrow,\uparrow\downarrow,\uparrow\downarrow,\uparrow\downarrow)$  is the basis where we expand $\epsilon_0,\,\epsilon_1$.  These are a basis of 32 component spinor, the first 2 entries correspond to the Lorentz group indices and the remaining 3 are the indices of the $\text{SO}(6)$ R-Symmetry so the supersymmetries of the defect are not mixed.

The total system is parametrized by the choices of $\epsilon_1$ that are also superconformal charges of the defect, this is the last 4 lines of the charges above.
\begin{eqnarray}
\label{eq:Def+WLsusy}
&&(\downarrow,\uparrow\downarrow,\uparrow,\uparrow,\uparrow).S\mp(\uparrow,\uparrow\downarrow,\downarrow,\downarrow,\downarrow).S\,,
\nonumber\\
&&(\downarrow,\uparrow\downarrow,\uparrow,\uparrow,\downarrow).S\pm (\uparrow,\uparrow\downarrow,\downarrow,\downarrow,\uparrow).S\,,
\nonumber\\
&&(\downarrow,\uparrow\downarrow,\uparrow,\downarrow,\uparrow).S\mp(\uparrow,\uparrow\downarrow,\downarrow,\uparrow,\downarrow).S\,,
\nonumber\\
&&(\downarrow,\uparrow\downarrow,\uparrow,\downarrow,\downarrow).S\pm (\uparrow,\uparrow\downarrow,\downarrow,\uparrow,\uparrow).S\,.
\end{eqnarray}
For each line and sign choice above the supersymmetry  that also preserves the Wilson loop is given by \eqref{eq:WLSCH}.

\subsection{String theory}

Now we move on to the supersymmetry preserved by the configuration in the gravity side. From the string theory point of view, supersymmetry transformations are parametrized by a Killing spinor $\epsilon$ which is a solution of equation dictated by the vanishing of gravitino variation. For the supergravity solution corresponding to AdS$_5\times$S$^5$, this equation takes the form
\be
\nabla_m\epsilon-\frac12\gamma\gamma_4\Gamma_m\epsilon=0\,,
\ee
where $\gamma=i\gamma_{0123}$ with $\gamma_i$ are 10d flat space Dirac matrices. On the other side, being $E^i_m$ the corresponding vielbein, we have the curved space Dirac matrices $\Gamma_m=E^i_m\gamma_i$. Solution of this equation can be written in the following form
\be
\epsilon(x)=\frac{e^{\frac{\phi}{2}\gamma_{12}}}{\sqrt{y}}H(\theta_a)\left(\epsilon_{-} + y \epsilon_{+} + t\gamma_{04}\epsilon_{+} + x_3\gamma_{34}\epsilon_{+}+re^{-\phi\gamma_{12}}\gamma_{14}\epsilon_{+}\right)\,,
\label{Killingesp}
\ee
where $\epsilon_{\pm}$ have positive/negative chirality with respect to $\gamma$ and therefore can be parametrized by two real spinors $\eta_1$ and $\eta_2$
\be
\epsilon_{+}=\left(1+\gamma\right)\eta_1 \qquad \epsilon_{-}=\left(1-\gamma\right)\eta_2\,,
\ee
and $H(\theta_a)$ is the solution of the internal space equation. For our particular solution \eqref{Killingesp} takes the form
\be
\epsilon(x)=\frac{e^{\frac{\phi}{2}\gamma_{12}}}{\sqrt{y}}h(\theta)\left(\epsilon_{-} + y \epsilon_{+} +  x_3\gamma_{34}\epsilon_{+}+re^{-\phi\gamma_{12}}\gamma_{14}\epsilon_{+}\right)\,,
\label{Killing}
\ee
and
\be
h(\theta)=e^{\frac{\theta}{2}\gamma\gamma_{45}}\,.
\ee
Charges preserved by a given configuration satisfy the kappa symmetry equation
\be
\left(1-\Gamma\right)\epsilon=0\,,
\label{kappa}
\ee
with the corresponding kappa symmetry projector
\be
\Gamma=\frac{\epsilon^{\alpha\beta}\partial_{\alpha}X^{m}\partial_{\beta}X^{n}}{2\sqrt{g}}\Gamma_{mn}K\,,
\label{proj}
\ee
with $K$ the corresponding conjugation operator.\footnote{Note that under our conventions, we can take a real representation of Dirac matrices subjected to the action
\be
K\epsilon_{\pm}=\pm\epsilon^{*}_{\pm} \qquad K\epsilon^{*}_{\pm}=\mp \epsilon_{\pm}\,.
\nonumber
\ee
With this definitions the kappa symmetry projector \eqref{proj} satisfies the properties required, namely $\text{tr}\Gamma=0$ and $\Gamma^2=1$.}
Introducing \eqref{Killing} in \eqref{kappa} and multiplying by $\sqrt{y}e^{-\frac{\phi}{2}\gamma_12}$ we obtain the following equation
\begin{align}
\begin{split}
\left(e^{-\phi\gamma_{12}}\tilde{\Gamma}-r'\gamma_{12}\right)h^{-1}(\theta)\left(-\epsilon^{*}_{-}+y\epsilon^*_{+}+\right. &\left. x_3\gamma_{34}\epsilon^{*}_{+}+re^{-\phi\gamma_{12}}\gamma_{14}\epsilon^{*}_{+}\right)=\\
&=rh(\theta)\left(\epsilon_{-}+y\epsilon_{+}+x_3\gamma_{34}\epsilon_{+}+re^{-\phi\gamma_{12}}\gamma_{14}\epsilon_{+}\right)\,,
\end{split}
\label{kappaeq}
\end{align}
where
\be
\tilde{\Gamma}=\left(y'\gamma_{24}+x'_3\gamma_{23}+my\gamma_{25}\right)\Rightarrow \tilde{\Gamma}^2=r'^2-r^2\,.
\ee
Time dependent terms have to vanish independently leading to the following relation
\begin{align}
\begin{split}
\epsilon^{*}_{-}=\left(y-\frac{rr'y'}{(r')^2-r^2}\right)\epsilon^{*}_{+}+\left(x_3+c\frac{rr'y^2}{(r')^2-r^2}\right)&\gamma_{34}\epsilon^{*}_{+}+m\frac{rr'y}{(r')^2-r^2}\gamma_{45}\epsilon^{*}_{+}\\
&-\frac{r^2}{(r')^2-r^2}\left(y'-cy^2\gamma_{34}-my\gamma_{45}\right)\gamma_{12}e^{\theta\gamma\gamma_{45}}\epsilon_{+}\,.
\end{split}
\label{relation}
\end{align}
A straightforward computation shows that $\tau$-independent part of \eqref{kappaeq} leads to the same relation.
Note that left hand side of \eqref{relation} is $\sigma$-independent, then consistency implies the right hand side to be so.
In the $\kappa\to\infty$ limit we find the following constraint
\be
\epsilon^*_{-}=R\cos\chi\gamma_{12}\epsilon_{+}+R\sin\chi\gamma\gamma_{1245}\epsilon_{+}+L\gamma_{34}\epsilon^*_{+}\,,
\ee
On the other hand, kappa symmetry equation for the D5-brane embedding leads to the additional condition \cite{CorreaYoung,Skenderis}
\be
\frac12\left(1+\gamma_{3456}\right)\epsilon=\epsilon\,.
\label{brane}
\ee
Note that both conditions are not compatible for arbitrary $\chi$, leaving only the $\chi=0$ case as the supersymmetric configuration.  This is in agreement with the gauge theory analysis of supersymmetry.


\begin{thebibliography}{99}

\bibitem{Karch:2000gx}
  A.~Karch and L.~Randall,
  JHEP {\bf 0106} (2001) 063
  doi:10.1088/1126-6708/2001/06/063
  [hep-th/0105132].

\bibitem{DeWolfe:2001pq}
  O.~DeWolfe, D.~Z.~Freedman and H.~Ooguri,
  Phys.\ Rev.\ D {\bf 66} (2002) 025009
  [hep-th/0111135].

\bibitem{Lee:2002cu}
  P.~Lee and J.~w.~Park,
  Phys.\ Rev.\ D {\bf 67}, 026002 (2003)
  doi:10.1103/PhysRevD.67.026002
  [hep-th/0203257].


\bibitem{DeWolfe:2004zt}
  O.~DeWolfe and N.~Mann,
  JHEP {\bf 0404}, 035 (2004)
  doi:10.1088/1126-6708/2004/04/035
  [hep-th/0401041].

\bibitem{Okamura:2005cj}
  K.~Okamura, Y.~Takayama and K.~Yoshida,
  JHEP {\bf 0601}, 112 (2006)
  doi:10.1088/1126-6708/2006/01/112
  [hep-th/0511139].

 \bibitem{Susaki:2005qn}
  Y.~Susaki, Y.~Takayama and K.~Yoshida,
  Phys.\ Lett.\ B {\bf 624}, 115 (2005)
  doi:10.1016/j.physletb.2005.07.058
  [hep-th/0504209].

\bibitem{Susaki:2004tg}
  Y.~Susaki, Y.~Takayama and K.~Yoshida,
  Phys.\ Rev.\ D {\bf 71}, 126006 (2005)
  doi:10.1103/PhysRevD.71.126006
  [hep-th/0410139].


\bibitem{Diaconescu:1996rk}
  D.~E.~Diaconescu,
  Nucl.\ Phys.\ B {\bf 503} (1997) 220
  doi:10.1016/S0550-3213(97)00438-0
  [hep-th/9608163].

  \bibitem{Giveon:1998sr}
  A.~Giveon and D.~Kutasov,
  Rev.\ Mod.\ Phys.\  {\bf 71} (1999) 983
  doi:10.1103/RevModPhys.71.983
  [hep-th/9802067].

  \bibitem{Constable:1999ac}
  N.~R.~Constable, R.~C.~Myers and O.~Tafjord,
  Phys.\ Rev.\ D {\bf 61} (2000) 106009
  doi:10.1103/PhysRevD.61.106009
  [hep-th/9911136].


\bibitem{Kristjansen:2012tn}
  C.~Kristjansen, G.~W.~Semenoff and D.~Young,
  JHEP {\bf 1301}, 117 (2013)
  doi:10.1007/JHEP01(2013)117
  [arXiv:1210.7015 [hep-th]].


\bibitem{Kristonepoint}
  I.~Buhl-Mortensen, M.~de Leeuw, A.~C.~Ipsen, C.~Kristjansen and M.~Wilhelm,
  Phys.\ Rev.\ Lett.\  {\bf 117}, no. 23, 231603 (2016)
  doi:10.1103/PhysRevLett.117.231603
  [arXiv:1606.01886 [hep-th]].

\bibitem{Buhl-Mortensen:2016jqo}
  I.~Buhl-Mortensen, M.~de Leeuw, A.~C.~Ipsen, C.~Kristjansen and M.~Wilhelm,
  arXiv:1611.04603 [hep-th].


\bibitem{deLeeuw:2015hxa}
  M.~de Leeuw, C.~Kristjansen and K.~Zarembo,
  JHEP {\bf 1508}, 098 (2015)
  doi:10.1007/JHEP08(2015)098
  [arXiv:1506.06958 [hep-th]].

\bibitem{Buhl-Mortensen:2015gfd}
  I.~Buhl-Mortensen, M.~de Leeuw, C.~Kristjansen and K.~Zarembo,
  JHEP {\bf 1602}, 052 (2016)
  doi:10.1007/JHEP02(2016)052
  [arXiv:1512.02532 [hep-th]].

\bibitem{deLeeuw:2016umh}
  M.~de Leeuw, C.~Kristjansen and S.~Mori,
  Phys.\ Lett.\ B {\bf 763}, 197 (2016)
  doi:10.1016/j.physletb.2016.10.044
  [arXiv:1607.03123 [hep-th]].

\bibitem{KristWL}
  M.~de Leeuw, A.~C.~Ipsen, C.~Kristjansen and M.~Wilhelm,
  arXiv:1608.04754 [hep-th].

\bibitem{Nagasaki}
  K.~Nagasaki, H.~Tanida and S.~Yamaguchi,
  JHEP {\bf 1201}, 139 (2012)
  doi:10.1007/JHEP01(2012)139
  [arXiv:1109.1927 [hep-th]].


\bibitem{Nagasaki:2012re}
  K.~Nagasaki and S.~Yamaguchi,
  Phys.\ Rev.\ D {\bf 86}, 086004 (2012)
  doi:10.1103/PhysRevD.86.086004
  [arXiv:1205.1674 [hep-th]].

\bibitem{chen}
B.~Chen, X.~J.~Wang and Y.~S.~Wu,
JHEP {\bf 0402}, 029 (2004)
[hep-th/0401016]



\bibitem{CorreaYoung}
  D.~H.~Correa and C.~A.~S.~Young,
  J.\ Phys.\ A {\bf 41}, 455401 (2008)
  doi:10.1088/1751-8113/41/45/455401
  [arXiv:0808.0452 [hep-th]].

\bibitem{Skenderis}
  K.~Skenderis and M.~Taylor,
  JHEP {\bf 0206}, 025 (2002)
  doi:10.1088/1126-6708/2002/06/025
  [hep-th/0204054].

\bibitem{Drukker:1999zq}
  N.~Drukker, D.~J.~Gross and H.~Ooguri,
  Phys.\ Rev.\ D {\bf 60}, 125006 (1999)
  doi:10.1103/PhysRevD.60.125006
  [hep-th/9904191].

\bibitem{Zarembo:2002an}
  K.~Zarembo,
  Nucl.\ Phys.\ B {\bf 643}, 157 (2002)
  doi:10.1016/S0550-3213(02)00693-4
  [hep-th/0205160].

\bibitem{Nahm:1977tg}
  W.~Nahm,
  Nucl.\ Phys.\ B {\bf 135}, 149 (1978).
  doi:10.1016/0550-3213(78)90218-3




  %
\bibitem{Pestun:2007rz}
  V.~Pestun,
  Commun.\ Math.\ Phys.\  {\bf 313}, 71 (2012)
  doi:10.1007/s00220-012-1485-0
  [arXiv:0712.2824 [hep-th]].

\bibitem{Drukker:2010jp}
  N.~Drukker, D.~Gaiotto and J.~Gomis,
  JHEP {\bf 1106}, 025 (2011)
  doi:10.1007/JHEP06(2011)025
  [arXiv:1003.1112 [hep-th]].



\end{thebibliography}
\end{document}